\documentclass{aip-cp}

\usepackage[numbers]{natbib}
\usepackage{rotating}
\usepackage{graphicx}
\usepackage{subcaption}

\usepackage{lineno}

\begin{document}
\title{Building Medium Size Telescope Structures for the Cherenkov Telescope Array}

\author[aff1]{A. Schulz\corref{cor1}}
\author[aff1]{M. Garczarczyk}
\author[aff2]{L. Oakes }
\author[aff1]{S. Schlenstedt}
\author[aff2]{U. Schwanke }
\author[aff3]{the CTA Consortium}

\affil[aff1]{DESY, Platanenallee 6, D-15738 Zeuthen, Germany}
\affil[aff2]{Humboldt University Berlin, Newtonstr. 15, D-12489 Berlin, Germany}
\affil[aff3]{See www.cta-observatory.org for full author \& affiliation list}
\corresp[cor1]{Corresponding author: anneli.schulz@desy.de}

\maketitle

\begin{abstract}
The Cherenkov Telescope Array (CTA) is the future instrument in ground-based gamma-ray astronomy in the energy range from 20\,GeV to 300\,TeV. Its sensitivity will surpass that of current generation experiments by a factor $\sim$10, facilitated by telescopes of three sizes. The performance in the core energy regime will be dominated by Medium Size Telescopes (MST) with a reflector of 12\,m diameter. A full-size mechanical prototype of the telescope structure has been constructed in Berlin. The performance of the prototype is being evaluated and optimisations, among others, facilitating the assembly procedure and mass production possibilities are being implemented. We present the current status of the developments from prototyping towards pre-production telescopes, which will be deployed at the final site.    

\end{abstract}	

\section{INTRODUCTION}

\begin{figure}%
  \centering
	\begin{tabular}{ p{0.4\textwidth} p{0.4\textwidth} }
		\includegraphics[height=0.3\textheight]{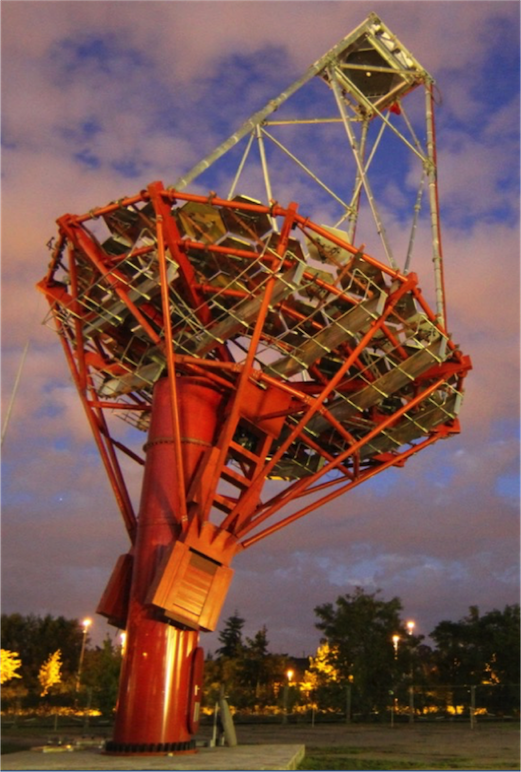} 
		\put(-11,4){a)}
		&
		\includegraphics[height=0.3\textheight]{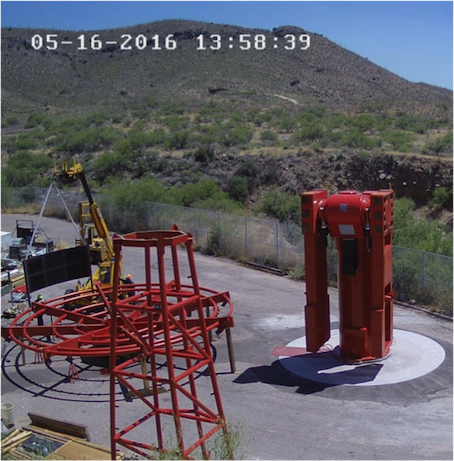}		
		\put(-11,4){b)}
	\end{tabular}
  \caption{Prototype of a) the MST structure in Berlin Adlershof and b) the Schwarzschild-Couder telescope in Arizona . }%
  \label{fig:prototyp}%
\end{figure}

The future instrument in very-high-energy (VHE, E$\geq $0.1\,TeV) gamma-ray astronomy, the Cherenkov Telescope Array (CTA), is presented in many contributions during the Gamma 2016. The overall status of the project is presented by \citet{hofmann_cta}. The physics topics which will be covered by the so-called key science projects, performed in the first years of operation by the CTA consortium, are presented by \citet{vercellone_ksp}. Overviews about the different telescope types are given, including the overall status of the medium size telescope (MST) presented by \citet{MST_puehlhofer}. MST includes three sub-projects, two solutions for the camera and the structure. The status of the NectarCAM is presented by \citet{glicenstein_nectar}, a test facility for FlashCam by \citet{flash_eisenkolb}. \\  
The structure of the MST is subdivided into the following parts: The mechanical assembly includes the positioner (made of a tower, head and yokes), the optical support structure (dish, counter weights, mirror access and camera support structure), the drive assemblies (for azimuth and elevation), the camera maintenance structure and the foundations. Mirrors, active mirror control (so-called actuators) and the mirror support unit form the optical assembly. The auxiliary assemblies include the electrical installations and power, telescope calibration, safety systems and software. \\   
This contribution focuses on the the developments by the MST structure team in the phase from prototyping to pre-production. A prototype for the MST structure was built in Berlin Adlershof in 2012, see Figure\,\ref{fig:prototyp} a), and used for extensive testing during the last years. In 2016 a prototype for the Schwarzschild-Couder telescope was constructed in Arizona. For this prototype the same positioner including the drive system as for the MST is used. Concerning the positioner this can be counted as the second prototype for the MST, too. New developments and redesigns in several sub-systems are presented in the following sections. \\
Several topics and developments within MST structure are covered by dedicated contributions, these are the calibration developments by \citet{calib_oakes}, the pointing solution by \citet{pointing_tiziani} and developments of the actuators by \citet{amc_diebold}. \\

\section{Head and Yokes}
For the prototype in Berlin the head and yokes were welded. To allow for a better force flow and to reduce costs these parts are casted for the second prototype, realised in Arizona, see Fig.\,\ref{fig:prototyp} b). The comprised improvements include also stow pins inside the head and improved ventilation and cabling systems. 

\section{Dish Optimization}
\begin{figure}%
  \centering
	\begin{tabular}{ p{0.4\textwidth} p{0.4\textwidth} }
	\includegraphics[width=0.35\textwidth]{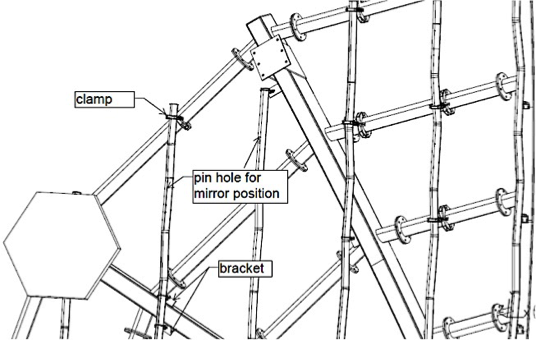} 
	\put(-160,110){a)}
	&
	\includegraphics[width=0.35\textwidth]{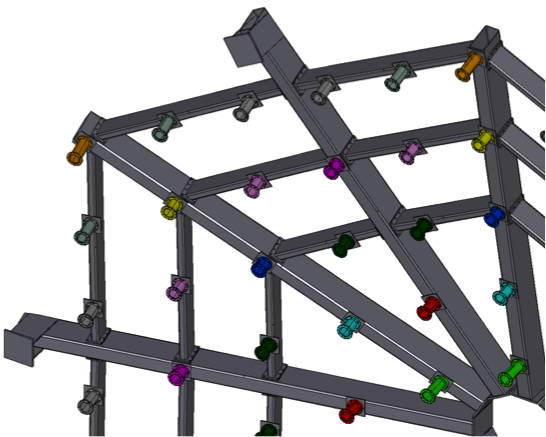}
	\put(-160,110){b)}	
	\end{tabular}
  \caption{Technical drawings of part of the dish before (a) and after (b) the optimisation, respectively. }%
  \label{fig:dish}%
\end{figure}

One part of the MST that was redesigned after the experiences with the construction of the prototype is the dish. The first realisation included an intermediate structure between the mirror mounts and the central beams, see Fig.\,\ref{fig:dish} a). This lead to a higher manpower need during the installation and had the negative effect that not every mirror was supported directly by the beams. The new design comprises a unified mirror interface, straight profiles and a weight reduced by 3 tons. The technical drawing of a part of the dish is shown in Fig.\ref{fig:dish} b). The new dish will be mounted in Berlin in autumn 2016 and tested during the following months.

\section{Drive System}
\begin{figure}%
  \centering
	\begin{tabular}{ p{0.4\textwidth} p{0.4\textwidth} }
	\includegraphics[width=0.4\textwidth]{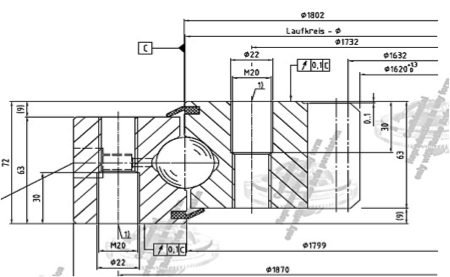} 
	\put(-160,110){a)}	
	&
	\includegraphics[width=0.35\textwidth]{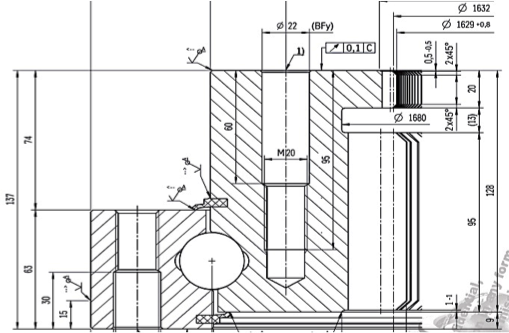}
	\put(-160,110){b)}
	\end{tabular}
  \caption{Azimuth bearing before a) and after b) the optimisation, respectively. }%
  \label{fig:drive}%
\end{figure}

The evaluation of the prototype resulted in a new azimuth bearing, old and new versions shown in Fig.\,\ref{fig:drive} a) and b), respectively. The new bearing allows for a higher torque safety margin and features a separated encoder section. The separation has the advantage of a smaller pitch and results in a higher position accuracy; the encoder is free of grease in the new solution. 

\section{Continuous Drive Tests}
\begin{figure}
  \centering
	\begin{tabular}{ p{0.5\textwidth} p{0.4\textwidth} }
	\includegraphics[width=0.45\textwidth]{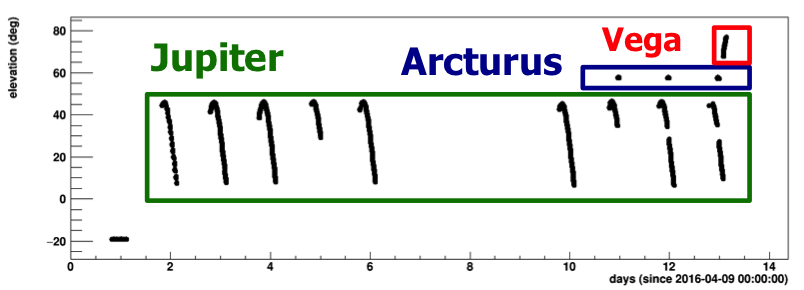} 
	\put(0,4){a)}
	&
	\includegraphics[width=0.35\textwidth]{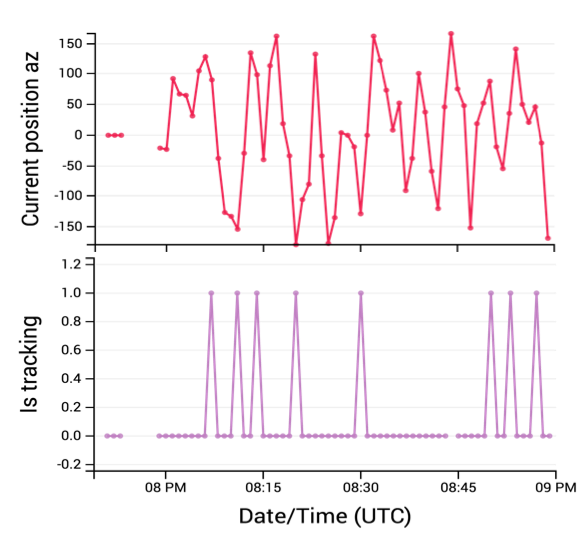}
	\put(-11,4){b)}
	\end{tabular}
  \caption{Examples of drive test results, see text for details.}%
  \label{fig:drivetest}%
\end{figure}

During the years continuous drive tests were preformed to monitor the system and test the software. Figure\,\ref{fig:drivetest} a) shows two weeks of tracking tests which included the tracking of Jupiter, Arcturus and Vega spanning over the entire elevation range. One hour of read out statuses of the drive system are shown in Figure\,\ref{fig:drivetest} b), the upper panel shows the current azimuthal position and the lower panel the status of "is tracking". Plots like these show the correct behaviour of the drive system and possible software issues. 

\section{Mirror Tests}
\begin{figure}%
  \centering
	\begin{tabular}{ p{0.3\textwidth} p{0.3\textwidth} p{0.3\textwidth} }
	\includegraphics[width=0.3\textwidth]{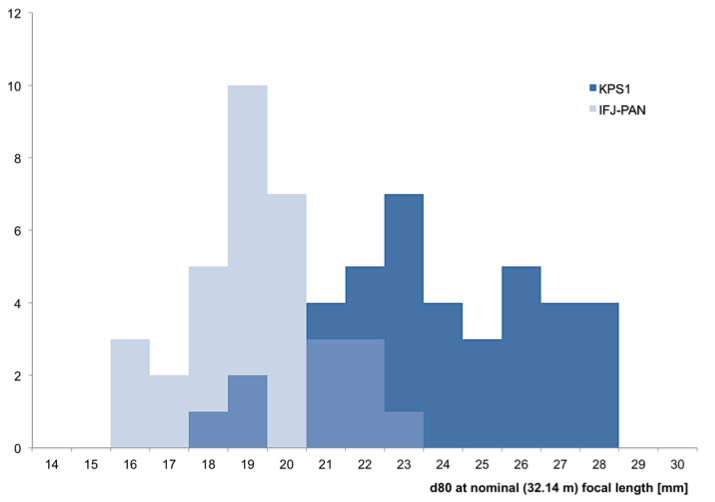} 
	\put(0,4){a)}
	&
	\includegraphics[width=0.3\textwidth]{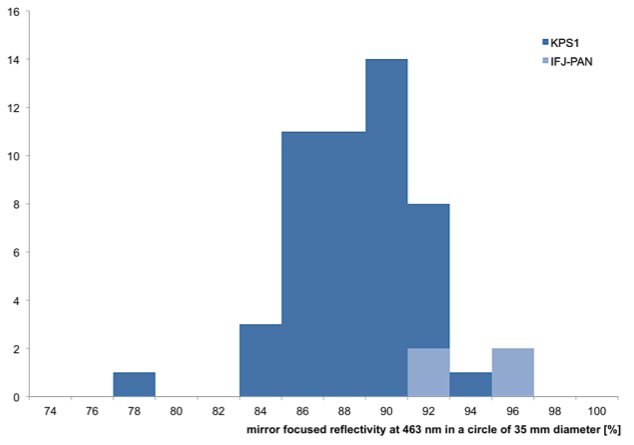}
	\put(0,4){b)}
	&
	\includegraphics[width=0.3\textwidth]{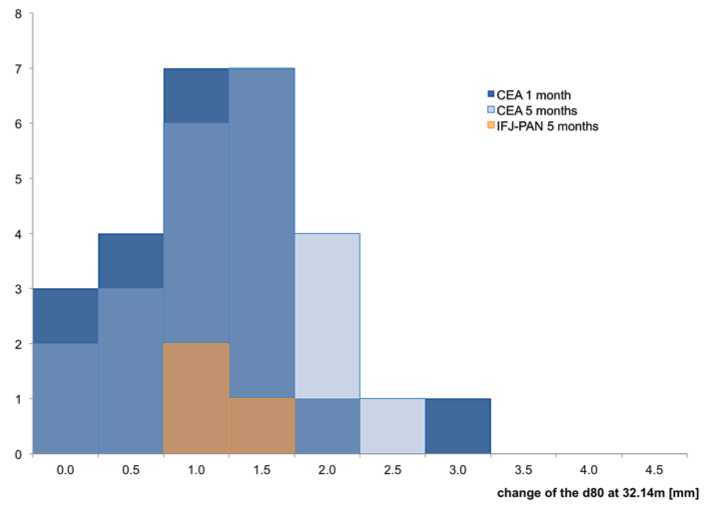}
	\put(0,4){c)}
	\end{tabular}
  \caption{Mirror test results: a) d80 at nominal focal length, b) focussed reflectivity, c) change of d80 after temperature cycling.   }%
  \label{fig:mirror}%
\end{figure}

Three teams produce mirror facets for the MST: INAF in Italy, CEA in France and IFJ-PAN in Poland with industrial partners. To ensure the quality of the individual mirror facets like the point spread function and reflectivity, continuous tests are performed. The point spread function is specified as the diameter of a circle around the spot mass center of gravity in the focal plane that contains 80\% of the total light intensity, thus denoted by d80. These tests include measurements after $>$300 temperature cycles between $-20^\circ - +30^\circ$ C, performed in a climate chamber at DESY and further tests with mirrors installed on the prototype. In Fig.\,\ref{fig:mirror} the test results of some mirrors are shown, the measurement of the point spread function and the reflectivity together with the change in point spread function after temperature cycling. No continuous worsening is seen for any of the mirrors. 

\section{Camera Interface Test}
\begin{figure}%
  \centering
	\includegraphics[width=0.35\textwidth]{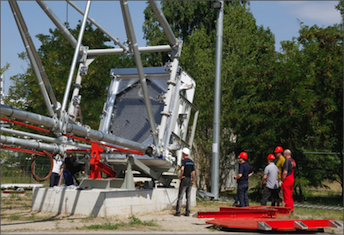} 
  \caption{Camera interface test performed together with the FlashCam team.  }%
  \label{fig:camera}%
\end{figure}

The interface to the camera was tested in June 2016, including the cooling system and the readout noise. The body of a full-size FlashCam prototype was mounted successfully in the prototype structure, as shown in Fig.\,\ref{fig:camera}. The concept of the LEDs used for the pointing system was investigated and optimised. 

\section{Quality and Documentation}
To prepare for the pre-production and production phase emphasis is also put on the preparation of quality assurance and documentation. In general the relevant parts of the ISO 9001 standard are followed and the Reliability, Availability, Maintainability and Safety (RAMS) data are standardised. An audit review process for the suppliers is foreseen as well as independent reviews of documents. 

\section{Schedule}
The prototype in Berlin Adlershof will be upgraded with the new dish in autumn 2016, more tests and requirement validations will follow. The MST team has to pass a pre-production readiness review before the telescopes can be installed on site. The current CTA schedule foresees the first telescopes of the pre-production to be installed in 2018 on the Southern site. The production will follow after another review ensuring production readiness. 

\section{CONCLUSION}
The recent developments show that the installation of the prototype and the testing of the systems was and is an important step towards pre-production. The prototype allowed for improvements concerning hardware and software as well as interface checks and lessons learned concerning the production and assembly processes.  

\section{ACKNOWLEDGMENTS}
We gratefully acknowledge support from the
agencies and organizations under Funding Agencies at www.cta-observatory.org. 


\bibliographystyle{aipnum-cp}%
\bibliography{references}%

\end{document}